\documentclass[useAMS,usenatbib]{mn2e}

\usepackage{epic}
\usepackage{epsfig}
\usepackage{rotating}
\usepackage{url}

\title[Possible signature of distant foreground in the Planck data]{Possible signature of 
distant foreground in the Planck data}
\author[V. N. Yershov, V. V. Orlov  and A. A. Raikov]{V. N. Yershov$^{1,2}$\thanks{E-mail:
v.yershov@ucl.ac.uk},
V. V. Orlov$^{2,3}$ and A. A. Raikov$^2$\\
$^1$Mullard Space Science Laboratory, University College London, 
Holbury St.Mary, Dorking, RH5 6NT U.K.\\
$^2$Main (Pulkovo) Astronomical Observatory 
of the Russian Academy of Sciences, 65 Pulkovskoe shosse, 
196140, St.Petersburg, Russia\\
$^3$Saint Petersburg State University, 
28 Universitetskij prospect Peterhof,  
198504, St.Petersburg, Russia}

\begin{document}

 \date{Accepted 2014 September 16. Received 2014 September 15; in original form 2013 December 17}

\maketitle

\begin{abstract}
By using the Planck map of the cosmic microwave background (CMB) radiation 
we have checked and confirmed the existence of a correlation 
between supernova (SN) redshifts, $z_{\rm SN}$, and CMB temperature fluctuations 
at the SNe locations, $T_{\rm SN}$, which we previously reported for the  Wilkinson 
Microwave Anisotropy Probe data. 
The Pearson correlation coefficient for the Planck data is $r=+0.38\pm 0.08$
which indicates that the correlation is statistically significant
(the signal is about $5\sigma$ above the noise level).
The correlation becomes even stronger for the type Ia subsample of SNe, 
 $r_{\rm Ia}=+0.45\pm 0.09$, whereas for the rest of the SNe 
it is vanishing. 
By checking the slopes of the regression lines $T_{\rm SN} / z_{\rm SN}$ 
for Planck's different frequency bands we have also excluded the possibility 
of this anomaly being caused by the Sunyaev-Zeldovich effect. 
The remaining possibility is some, 
unaccounted for, contribution to the CMB from distant ($z>0.3$) 
foreground through either the integrated Sachs-Wolfe effect or 
thermal emission from intergalactic matter. 
\end{abstract}

\begin{keywords}
supernovae:  general ---  galaxies: star formation --- dust ---
cosmic background radiation ---  methods: statistical
\end{keywords}

\section{Introduction}

Previously  we reported \citep{yershov12} an anomaly  found in the
cosmic microwave background (CMB) radiation, that manifested itself
as a positive correlation between supernova (SN) redshifts and 
CMB temperature fluctuations measured by the  
Wilkinson Microwave Anisotropy Probe (WMAP) at the SN locations.   
In particular, the WMAP CMB map temperatures
corresponding to different SN redshifts revealed an excess 
of temperatures $+29.9\pm4.4 ~[\mu{\rm K}]$ for the redshift range $z\sim$
0.5 to 1.0.

\begin{figure}
\hspace{-0.5cm}
\epsfig{figure=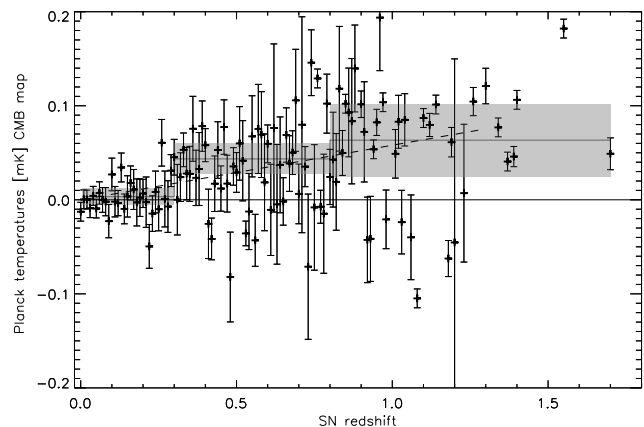,width=9.0cm}
\vspace{-0.4cm}
\caption{Planck CMB temperatures $T_{\rm SN}$ near
the supernovae locations as a function of SN redshift $z_{\rm SN}$. 
The dashed line  indicates the slope $\xi=+58.0\pm 9.0\, [{\mu}\rm K]$ of the 
linear regression $T_{\rm SN}=\xi\,z_{\rm SN}$. The shaded area 
indicates the $\pm3\sigma$ tolerances around the $\overline{T}_{\rm SN}$ averaged 
within three SN redshift intervals: $z<0.3$,
$0.3<z<0.8$ and $z>0.8$.} 
\label{fig:sn_temp_redshift}
\end{figure}

When discussing the possible origin of this anomaly,
we have excluded the energy-dependent Sunyaev-Zeldovich (SZ) effect
\citep{sunyaev70} by  comparing the magnitude 
of the anomaly in different WMAP frequency bands. 
The remaining possibilities were the contribution from either dust or 
the integrated Sachs-Wolfe (ISW) effect \citep{sachs67}. 
For example, a similar band-independent 
correlation between WMAP temperatures and large-scale structures 
traced by galaxies was found by \cite{goto12} for the Wide-field 
Infrared Survey Explorer data, which was attributed by 
the authors to the Sachs-Wolfe effect.
This effect alters the energy of CMB photons
when these photons traverse gravitational fields of different 
strengths corresponding to either voids or clumps of matter,
the photon energy being increased for clumps and decreased for voids.

In the case of SNe, both effects could be simultaneous because SNe obviously
occur in clumps of matter, but many of them are also known to be 
affected by high extinction in their host galaxies, indicating the presence of dust
\citep{dipaola02,vink04,elias08,kozlovski10,romero11,mattila12}.
   
After the release of the ESA's Planck space telescope data
\citep{planck14}, it was instructive to check the presence of the same   
anomaly using a higher quality data set obtained 
by a different instrument for different frequency bands. 
Here we present the results of such a check for the released 
Planck CMB map, as well as for its five (of nine) frequency bands.

\section{Planck CMB data and the supernova redshifts}

For our analysis we have taken the list of supernovae \citep{bartunov07} 
maintained by the Sternberng Astronomical Institute (Moscow) 
at \url{http://www.sai.msu.su/sn/sncat} 
which contains a large sample of SNe with known redshifts
(6365 SNe as for October 2013).
The Planck data products were obtained from ESA's Planck Legacy
Archive \citep{planck14} at \url{http://www.sciops.esa.int/pla/aio}, and the software for 
processing these data was adopted from the comprehensive
{\scriptsize HEALPIX} package developed by Jet Propulsion Laboratory and
available at \url{http://healpix.jpl.nasa.gov} \citep{gorski05}.

We have constructed a diagram of the 
Planck CMB pixel temperatures at the SN locations as a function 
of SN redshifts. Figure~\ref{fig:sn_temp_redshift} shows this diagram
for the Planck CMB map produced by the semi-parametric 
processing pipeline (SMICA R1.20). 

In order to avoid contamination from point sources and 
from the galactic plane, we have used  
the Planck confidence map containing pixels with an expected
low level of foreground contamination, as well as the Plank High
Frequency Instrument's (HFI) point source and galactic plane masks.
Additionally, we have restricted our SN sample to hight galactic latitudes
$|b\,|>40^\circ$ and to the SN redshifts $z_{\rm SN}>0.05$. The 1048 SNe with
$z<0.05$ were irrelevant for our study, so the remaining sample
used for checking the relationship between the Planck CMB map pixel temperatures
and SN redshifts consisted of 3348 SNe.  
By grouping the Planck CMB pixel temperatures into the SN redshift bins
with the bin size of 0.01, we have obtained the diagram 
shown in Fig.~\ref{fig:sn_temp_redshift}.

The average temperature
for the selected latitude range 
can differ from zero by being based on a subset
of the Planck data (the zero average corresponds to the whole 
set of pixels covering the entire celestial sphere). Since we are interested in 
the possible relationship between 
the SN redshifts and the CMB temperature 
fluctuations, we can subtract any constant value
from the bin-averaged temperatures. What is important is the slope $\xi$ of the regression
line $T_{\rm SN}=\xi\, z_{\rm SN}$ characterising the relationship between 
the SN redshifts and CMB temperatures. This regression line with $\xi=58.0 \pm 9.0$ 
(in ${\mu}\rm K$ per redshift unit) is shown in Fig.\ref{fig:sn_temp_redshift}
as the dashed line.
The temperature zero-level for this diagram was adjusted in such a way that to
nullify the average 
$\overline{T}_{\rm SN}^{0.05}=+2.7~[{\mu}\rm K]$  
for the local SNe with $z_{\rm SN}<0.05$. 

Like in the WMAP case, the diagram shown in  Fig.\ref{fig:sn_temp_redshift}
exhibits a positive anomaly of $T_{\rm SN}$ for high redshifts. 
The error-bars of $T_{\rm SN}$ in this diagram  
correspond to the weighted standard errors ($SE_w$) of the averages $\overline{T}^w_{\rm SN}$
in each redshift bin calculated as
\[
SE_w=\sqrt{\sigma_w^2  \sum_{i=1}^n w_i^2 / (\sum_{i=1}^n w_i)^2},
\]
where
\[\sigma_w^2 = \sum_{i=1}^n w_i(T_{\rm SN}^i-\overline{T}^w_{\rm SN})^2/ (\sum_{i=1}^nw_i-1),
\]
\[\overline{T}^w_{\rm SN}=\sum_{i=1}^n w_iT^i_{{\rm SN}}/ \sum_{i=1}^n w_i\,,\]
 and $n$ is the number of SNe in the bin. The individual weights $w_i$ of the 
Planck CMB pixel temperatures $T^i_{\rm SN}$ were obtained by using the
variances given for these temperatures in the Plank data tables. 
The points in Fig.\ref{fig:sn_temp_redshift} corresponding to the cases with 
a single source in a redshift bin (which are mostly the points with $z_{\rm SN}>1.3$)
are shown with the error-bars transferred from the Planck variance map for the 
CMB temperatures $T^i_{\rm SN}$.
The average number of SNe in each redshift bin 
was 60.8 for $z_{\rm SN}<0.5$ and 2.6 for $z_{\rm SN}\in(0.5,1.0)$, with the 
total number of 3348 SNe in the selected latitude range. 

Since the number of $T^i_{\rm SN}$ values in each bin is relatively 
small for high redshifts, we have also produced $\overline{T}_{\rm SN}$ 
averaged within three much larger
redshift intervals, $z_1<0.3$, $0.3<z_2<0.8$ and $0.8<z_3<1.7$, whose sizes
are roughly proportional to the physical distance intervals. The results 
of this averaging are presented in Table~\ref{bin3} and
in the form of shaded areas in 
Fig.\ref{fig:sn_temp_redshift} indicating 
the $\pm3\sigma$ tolerance intervals for the averages.
This three-bin averages highlight the anomaly more clearly,
showing that the $\overline{T}_{\rm SN}$ signal exceeds
the noise level by about $7\sigma$ for $z_2$.

\begin{table}
 \caption{Temperatures $\overline{T}_{\rm SN}$ averaged for three large 
redshift intervals (bins) 
 $0.0<z_1<0.3$, $0.3<z_2<0.8$ and $0.8<z_3<1.7$.}
 \label{bin3}
 \begin{tabular}{@{}llll}
  \hline
  {\scriptsize Characteristics} & $z_1$ & $z_2$ & $z_3$   \\
  \hline
  $\overline{T}_{\rm SN} ~[{\mu}K]$ &  $+3.2$ & $+43.5$ & $+63.4$ \\
  $\sigma_{\overline{T}_{\rm SN}} ~[{\mu}K]$ &  $\pm2.9$ & $~\pm5.4$ &  $\pm12.9$ \\
  Signal-to-noise ratio $~[{\sigma}]$  & ~~~1.0 & ~~~~7.9 &  ~~~~4.9 \\
    Sample size & ~2741  & ~~~~516 & ~~~~~91 \\
  \hline
 \end{tabular}
\end{table}

The magnitude of this anomaly is higher than the anomaly previously 
reported for the WMAP data, which can be attributed to the different from WMAP
set of frequency bands used in the Planck experiment. 
Part of this difference might also be due to the lesser averaging effect in the Planck 
data compared to WMAP because of higher angular resolution of the Planck instruments.

The Pearson correlation coefficient for our binned data is $r=0.38\pm 0.08$. The 
confidence interval here was calculated by bootstrapping the correlation \citep{efron79}.
This method consists in repeatedly resampling the data many times, 
each time calculating the correlation coefficient and, finally, 
getting the standard deviation of the generated in this way distribution of $r$ values. 
The calculated confidence interval does not include $r=0$ at about $4.6\sigma$ 
significance level, which indicates the statistical significance of the 
correlation. 
The same bootstrapping method was also used for calculating the 
tolerance interval ($\pm 9.0 ~[\mu {\rm K}]$) corresponding to
the slope of the regression line $T_{\rm SN} / z_{\rm SN}$ 
(the dashed line in Fig.\ref{fig:sn_temp_redshift}). 

The Pearson correlation coefficient calculated for the unbinned SN
data is essentially smaller, but so is its bootstrap confidence interval: 
$r_{\rm unbinned}=0.130\pm 0.016$. 
This shows that the estimate of the statistical significance
of the found correlation does not depend on the degree of data binning\,/\,smoothing.
Moreover, the calculation of the slope of the regression line 
for the unbinned data gives the same result as for the case when the
data are binned: $\xi_{\rm unbinned}=57.1 \pm 7.0 ~[\mu {\rm K}]$.

\begin{figure}
\hspace{-0.5cm}
\epsfig{figure=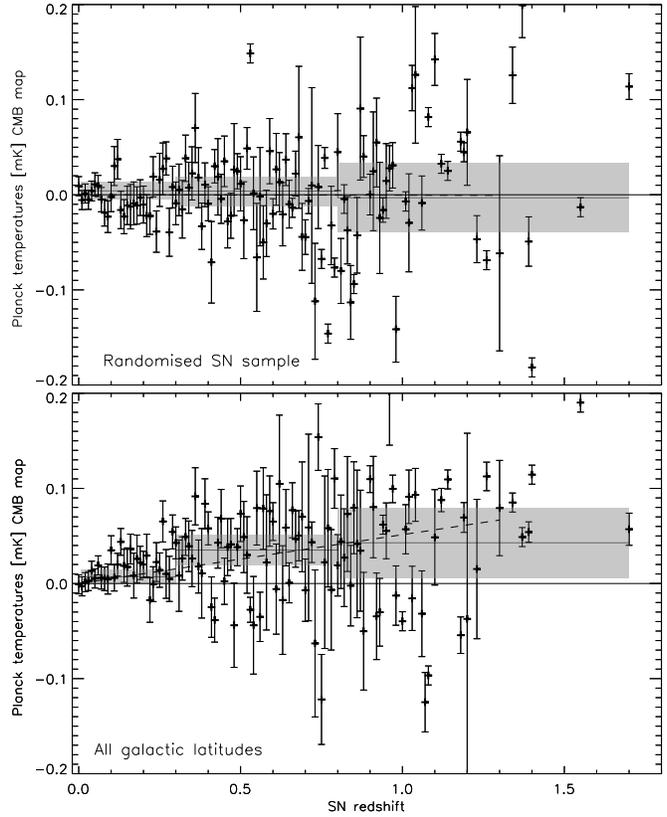,width=9.0cm}
\vspace{-0.2cm}
\caption{Upper panel: the $T_{\rm SN}/z_{\rm SN}$
diagram for the randomised SN sample does not show any
temperature anomalies; lower panel: the $T_{\rm SN}/z_{\rm SN}$
diagram for the whole SN sample including the SNe within the 
galactic plane shows a small but still statistically significant 
anomaly for the redshifts
$0.3<z_2<0.8$. The shaded areas indicate the $3\sigma$-tolerances 
around the $\overline{T}_{\rm SN}$ averaged 
for three SN redshift intervals: $z_1<0.3$,
$0.3<z_2<0.8$ and $z_3>0.8$.} 
\label{fig:sn_temp_rand}
\end{figure}

Yet another check ensuring that the seen correlation is not due to a systematic error
in our calculations consists in randomising the SN positions on the sky. For this purpose, 
we have used the same SNe but with their galactic longitudes randomised, while 
keeping the same declination range ($|b|>40^\circ$) as for the original SN sample.
The $T_{\rm SN}/z_{\rm SN}$ diagram for the randomised SN sample 
(shown in the upper panel of Fig.\ref{fig:sn_temp_rand}) does not reveal any 
temperature anomalies, the average
temperatures for the three selected redshift intervals being
$\overline{T}^{z_1}_{SN_{\rm rand}}=4.1\pm 3.1~[{\mu}\rm K]$,
$\overline{T}^{z_2}_{SN_{\rm rand}}=3.9\pm 5.1~[{\mu}\rm K]$,
$\overline{T}^{z_3}_{SN_{\rm rand}}=-3.1\pm 12.1~[{\mu}\rm K]$.

The latter test confirms that the positive anomaly seen in the Planck CMB pixel
temperatures is indeed related to the SN locations. Those, in turn,
trace the presence of distant ($z>0.3$) concentrations of matter 
(distant foreground), which might be behind the origin of the anomaly.    

Although the processing scheme used for generating the Planck CMB 
map was taking into account the local foreground emission 
from gas and dust belonging to our Galaxy, some fraction of this emission 
still might be left unaccounted for in the Planck CMB map product.  
It is for this reason that we have used a subsample of SNe,
limiting their coordinates to the region with the galactic latitudes 
$|b|>40^\circ$.
However, it would be instructive to check to what degree the local foreground,
obscures the systematic relationship between $T_{\rm SN}$ and $z_{\rm SN}$ seen for 
the high-latitude SNe. For this purpose we have performed the correlation analysis
of the Planck CMB map for the sample of 5318 SNe covering the whole galactic latitude range
(still the SNe with $z_{\rm SN}<0.05$ were excluded as irrelevant for our analysis).  
The lower panel of Fig.\ref{fig:sn_temp_rand} shows the $T_{\rm SN}/z_{\rm SN}$ 
diagram built for the whole SN sample, that includes also the SNe within
the galactic plane region. As expected for this data, the significance 
of the $T_{\rm SN}/z_{\rm SN}$ correlation is reduced,
with the Pearson correlation coefficient 
being $r=0.23\pm 0.10$ (as before, the confidence interval for this
coefficient was calculated by bootstrapping the correlation).

\begin{table}
 \caption{Temperatures $\overline{T}^{\rm gal}_{\rm SN}$ averaged for three large 
redshift intervals (bins) 
 $0.0<z_1<0.3$, $0.3<z_2<0.8$ and $0.8<z_3<1.7$ for the whole
SN sample, including the SNe near the galactic plane 
(see Fig.\ref{fig:sn_temp_rand} lower panel).}
 \label{bin3whole}
 \begin{tabular}{@{}llll}
  \hline
  {\scriptsize Characteristics} & $z_1$ & $z_2$ & $z_3$   \\
  \hline
  $\overline{T}^{\rm gal}_{\rm SN} ~[{\mu}{\rm K}]$ &  $+7.0$ & $+35.1$ & $+42.8$ \\
  $\sigma_{\overline{T}^{\rm gal}_{\rm SN}} ~[{\mu}{\rm K}]$ &  $\pm2.4$ & $~\pm5.3$ &  $\pm12.3$ \\
  Signal-to-noise ratio $~[{\sigma}]$  & ~~~2.8 & ~~~~6.6 &  ~~~~3.5 \\
    Sample size & ~4601  & ~~~~608 & ~~~109 \\
  \hline
 \end{tabular}
\end{table}

\begin{table*}
 \caption{Catalogue of supernovae stars with the associated Planck CMB map pixel 
temperatures at the SN locations (this table shows only the first and the last 
three entries of the catalogue, whereas the full catalogue is available online
in the form of an ASCII table).}
 \label{sncmbcat}
 \begin{tabular}{@{}cccclrrrrcccl}
  \hline
  N & SN &  R. A. &  Decl.   & ~~$z$ & $T_{\rm CMB}$ & $T_{\rm err}$ &
$l$~~ & $b$~~ & Galaxy &  Galaxy &  Mask &  SN \\
                & name &  $~~h~~m~~s$ & ~~~~~$^\circ$~~~$'$~~~$''$  &   & [$m$K] & [$m$K] &
[deg] & [deg] & name &  type &    & type  \\
  \hline
0001 & 1885A & 00 42 43.00 & +41 16 04.0 & 0.000 & -0.097 & 0.041 &  121.1698 & -21.5741 & NGC0224 &  Sb & 0 & IPec \\
0002 & 1938C & 13 16 08.28 & +25 09 40.0 & 0.000 &  0.122 & 0.023 &  14.7736  &  84.1190 &  -      &  -  & 0 & - \\
0003 & 1950D & 08 43 05.44 & +18 09 48.0 & 0.000 & -0.195 & 0.010 &  207.8540 &  32.5486 &  -      &  -  & 0 & - \\
\dots & \dots  & \dots &  \dots & \dots & \dots & \dots & \dots  & \dots &  \dots  & \dots  & \dots & \dots \\
6357 & 2002fx &  03 32 06.80 & -27 44 34.4 & 1.400 & 0.114 & 0.009 & 223.4417 & -54.5049 &  -      &  -  & 0 & Ia \\
6358 & 2003ak & 03 32 46.90  & -27 54 49.3 & 1.551 & 0.190 & 0.005 & 223.7699 & -54.3841 &  -      &  -  & 0 & Ia \\
6359 & 1997ff & 12 36 44.38  & +62 12 41.9 & 1.700 & 0.057 & 0.017 & 125.9055 &  54.8317 &  -      &  -  & 0 & - \\
  \hline
 \end{tabular}
\end{table*}

Nevertheless, the noisy effect of the local galactic foreground on the
$T_{\rm SN}/z_{\rm SN}$ correlation is not very big:
the average $\overline{T}_{\rm SN}$ values  calculated  for three large 
redshift bins still indicate a $6.6\sigma$-anomaly for the redshift
range  $0.3<z_2<0.8$. These $\overline{T}_{\rm SN}$ values for the whole 
sample, as well as their standard errors, 
signal-to-noise ratios and SN sample sizes, are given in Table~\ref{bin3whole}.
By comparing Tables \ref{bin3} and \ref{bin3whole} we can conclude that
the local galactic foreground might obscure the systematic contribution
from the distant foreground by about 30\%.
In order to facilitate further checks of the found anomaly 
by the other researches, we have generated the full catalogue of SNe 
with the associated Planck CMB map 
temperatures at the SN locations, which we have provided
in the form of an Appendix to this paper (an ASCII table available online). 
A small extract of this catalogue is presented in Table~\ref{sncmbcat} as 
an example showing the first three and the last three entries. 
This catalogue includes 6359 SN entries arranged in the 
ascending SN redshift order (6 entries were excluded as they were 
lacking the SN sky coordinates). The columns of this table provide the 
source number N, the source name (SN name), right ascension R.A. 
(in hours minutes and seconds), and declination Decl. (in degrees, arcminutes and arcseconds)
of the source, its redshift $z$, Planck CMB map pixel temperature at the SN location 
$T_{\rm CMB}$ and its variance $T_{\rm err}$ (both in $m$K), the SN 
galactic longitude $l$ and latitude $b$ (both in degrees), the name
and type of the SN host galaxy (if available), the value of the combined
galactic plane and point source mask and the SN type (if available).

\section{Distinction between different supernova types}

As we have mentioned, one of our interpretations of the revealed
anomaly was that it might be due to some possible  
contribution of thermal emission from dust residing in 
 the SN host galaxies. 
For checking this possibility, we have split our SN sample 
into two subsamples, one containing only the SNe of type Ia
(1937 sources for $|b\,|>40^\circ$), and 
the other with the rest of the SNe (1411 sources for $|b\,|>40^\circ$, 
mostly of the types II, Ib and Ic). 
The SNe from the latter sample   
typically occur in galaxies with higher star formation rates, 
while SNe Ia can occur in any galaxy \citep{filippenko89,farrah02,childress13}. 
What we expected was an enhancement of this correlation for the SNe 
in the galaxies with high star formation rates and containing 
large amounts of dust, that is, for the type II, Ib and Ic SNe. 
 
However, contrary to what we expected, the significance of the correlation 
became higher for the Ia type SNe, 
with the correlation significance $p$-value being reduced from $2.9\cdot 10^{-5}$
for the whole SNe sample to $7.3\cdot 10^{-9}$ for the SN of the type Ia subsample, 
whereas for the rest of the SNe the $p$-value increased dramatically to
$p=0.812$ and the significance of the Pearson's correlation 
coefficient decreased to $0.9\sigma$, revealing no correlation whatsoever for the 
other SN types. 
The results of our statistical tests for these two SN subsamples,
as well as for the whole sample, are summarised in Table \ref{slopes2}. 
The last entry in this Table gives the number of sources per sample in the selected 
latitude range.
The first entry gives the slopes $\xi$ of the regression 
lines characterising the relationship between $T_{\rm SN}$ 
and $z_{\rm SN}$ expressed in  $\mu {\rm K}$ per unit redshift interval.
The steepest of these slopes corresponds to the type Ia SNe.
The confidence intervals for the Pearson's correlation coefficients
(second line of the Table) were calculated by bootstrapping the correlation.
The correlation significance $p$-value is given in the third line of this Table. 
The lowest $p$-value, which indicates the highest statistical significance 
of the correlation, 
also points at the type Ia SNe as the objects likely to be responsible 
for the correlation. Therefore, our next statistical test (Section 4)
was made by using only the sample of the type Ia SNe.     

\begin{table}
 \caption{Results of the correlation tests for three SN samples.}
 \label{slopes2}
 \begin{tabular}{@{}llll}
  \hline
  {\scriptsize Characteristics} & All SNe & SN Ia & The rest of SNe   \\
  \hline
  $\xi ~[{\mu}{\rm K}]$ &  $58.0\pm 9.0$ & $72.2\pm 10.9$ & ~~$20.7\pm 24.1 $ \\
  Pearson's $r$ &  $0.38\pm 0.08$ & $0.45\pm 0.09$ &  $-0.13 \pm 0.14$ \\
  p-value &     $2.9\cdot 10^{-5}$ & $~7.3\cdot 10^{-9}$ & ~~~0.812 \\
  $\overline{T}_{\rm SN}^{(0.8,1.7)} [{\mu}{\rm K}]$ & $+63.4\pm 12.9$ & $76.6 \pm 11.8$ & $-36.9 \pm 34.9$ \\
 Sample size     &  ~~~~3348   &  ~~~~1937   &  ~~~~1411   \\
  \hline
 \end{tabular}
\end{table}

\section{Correlation tests for different Planck frequency bands}

When analysing the anomaly in the WMAP data, we have
checked (and rejected) the possibility of this anomaly being caused by the 
Sunyaev-Zeldovich effect. There were only small, statistically insignificant,
variations in the slopes of the regression lines $T_{\rm SN}/z_{\rm SN}$ obtained 
for three WMAP frequency bands, 40, 60 and 90 GHz. In fact, it was 
difficult to expect detecting the SZ effect by using the WMAP data 
because in order to see this effect clearly one has to use frequency bands
below and above the so-called cross-over 
frequency 218~GHz, whereas all three WMAP bands which we used for this check  
had frequencies below 218 GHz. 
   
The HFI and Low Frequency Instrument (LFI) on-board the Planck satellite provided a much wider
frequency coverage by Planck's nine frequency bands ranging from 30 to 857 GHz.
Although the Planck legacy archive contains just one CMB map generated
by combining the measurements made in the HFI frequency bands, 
these bands contain information about the foreground emission 
which is a mixture of the local and remote 
foreground emissions, with a larger contribution from the 
local foreground at low galactic latitudes and lower frequencies. 
Actually, combinations of different frequency bands
 were used by the Planck team for producing the Planck 
foreground emission maps. 
We can use both the frequency band maps and 
foreground emission maps for our correlation analysis, 
provided that we apply 
the Planck point source and galactic plane masks and exclude 
the galactic latitudes $|b|<40^\circ$ from the consideration
to reduce the noisy effect from the local foreground.

\begin{figure}
\epsfig{figure=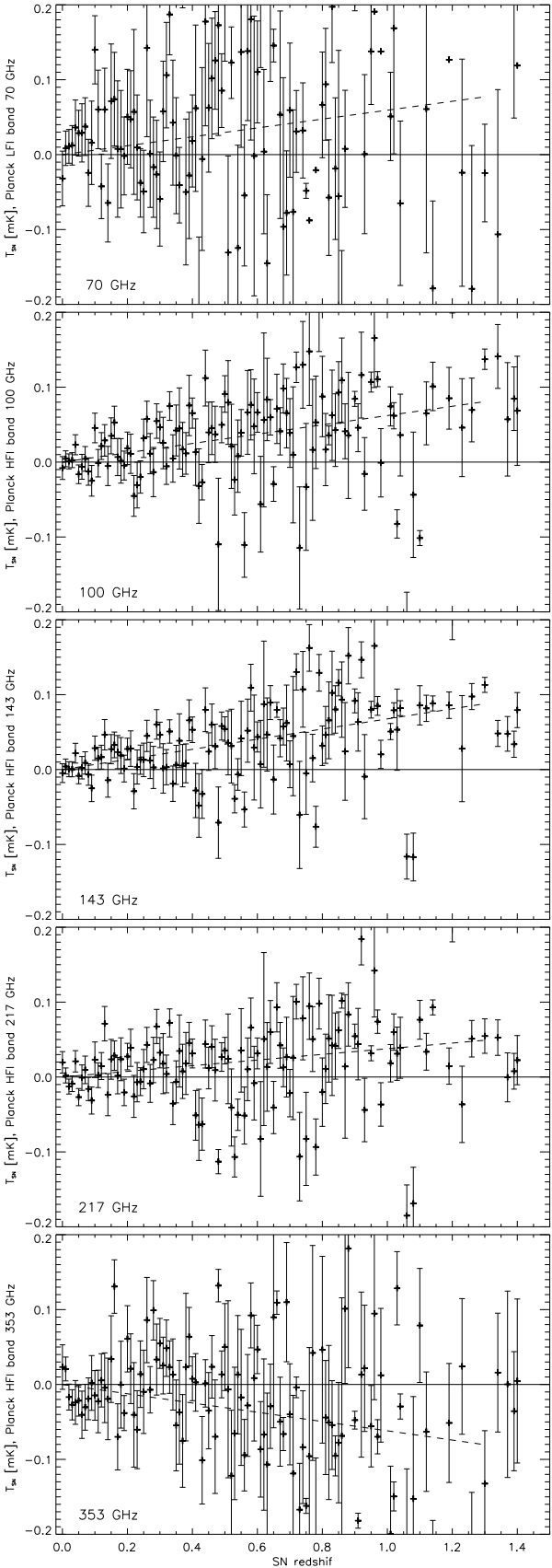,width=8.2cm}
\caption{$T_{\rm SN}/z_{\rm SN}$ diagrams constructed for five Plank 
frequency bands: LFI 70 GHz, HFI 100, 143, 217 and 353 GHz.} 
\label{fig:sn_temp_panels}
\end{figure}

When building the diagrams for different frequency bands,
 we have found that the scatter of the binned 
temperatures $T_{\rm SN}$ was extremely large for the bands 
30, 44, 545 and 857 GHz, so we have 
used only the remaining five Planck frequency bands 70, 100, 143, 217 and 353 GHz 
for which the scatter of the data points was tolerable
(these diagrams are shown in Fig.\ref{fig:sn_temp_panels}).
The results of our correlation tests applied to these five
diagrams are presented in Table \ref{slopes1}. 

The lowest significance of the correlation was obtained for 
the Planck LFI 70 GHz frequency band. For this band $r=0.12\pm0.09$, 
which means that the null-hypothesis about the absence
of any correlation in this case cannot be rejected.
However, the slopes of the regression lines $T_{\rm SN} / z_{\rm SN}$ 
for the Planck bands 70 and 100 GHz, $\xi=+59.3 \pm 36.4 \mu{\rm K}$ 
and $\xi=+62.1 \pm 13.6 \mu{\rm K}$, match, within the 
$3\sigma$ tolerance intervals, the slopes corresponding to the WMAP 
bands of 61 and 94 GHz, $\xi=+36.6 \pm 7.6 \mu{\rm K}$ and 
$\xi=+32.2 \pm 7.8 \mu{\rm K}$, respectively, 
which were obtained in our previous work \citep{yershov12}. 
In the Planck case the low significance of these 
coefficients is due to the fact that they
were calculated by using frequency maps containing 
emission from local foreground, whereas in the WMAP case
it was possible to use cleaned CMB maps.    
The highest significance of the correlation is obtained for the 
HFI 143 GHz frequency band with the significance level of $5.1\sigma$ 
for the correlation coefficient. 
 
The Sunyaev-Zeldovich effect should cause a decrease in the CMB intensity at 
the frequencies below 218 GHz and an increase at higher frequencies.
Therefore, if the observed anomaly were caused by this effect,
we would expect a higher positive anomaly in our plots for the 353 GHz band.
However, the effect we see for this band 
in Fig.\ref{fig:sn_temp_panels} (lower panel) is the opposite: the anomaly 
in the 353 GHz band is negative, and the slope of the 
regression line for this band is negative as well, 
$\xi_{353}=-61.8 \pm 18.2~[\mu{\rm K}]$.
Thus, we can confidently exclude the SZ effect as the possible 
cause of the discussed anomaly. 

\begin{table}
 \caption{Slopes of the regression lines characterising the relationship between 
$T_{\rm SN}$ and $z_{\rm SN}$ for five Planck frequency bands, two
foreground component maps (the low-frequency foreground map LFF, 
thermal dust foreground, TDF) and for the CMB lensing potential
map CLPM.}
 \label{slopes1}
 \begin{tabular}{@{}lrcc}
  \hline
   Planck       & Slope $\xi_\nu$~~~ & Pearson's   & Significance \\
   band           & $[{\mu}\rm K]$~~~~~          &  $r~~$   & $[\sigma]$  \\
  \hline
  ~70 {\scriptsize GHz}   & $59.3\pm 36.4$   & $0.12\pm 0.09$ & 1.3 \\
  100 {\scriptsize GHz}   & $62.1\pm 13.6$   &  $0.35\pm 0.11$ & 3.1 \\
  143 {\scriptsize GHz}  & $67.6\pm 6.3$   & $0.43\pm 0.08$  & 5.4 \\
  217 {\scriptsize GHz}  & $37.7\pm 8.4$   & $0.21\pm 0.09$  &  2.3 \\
  353 {\scriptsize GHz}  & $-61.8\pm 18.2$  & $-0.27\pm 0.09$ & 3.0 \\
\hline
LFF ~30 {\scriptsize GHz} & $-52.7\pm 15.0$ & $-0.26\pm 0.12$ & 2.2 \\
TDF 353 {\scriptsize GHz} & $-73.1\pm 29.0$ & $-0.34\pm 0.08$ & 4.2 \\
CLPM & $-55.5\pm 61.1$ & $-0.12\pm 0.11$ & 1.1 \\
\hline
 \end{tabular}
\end{table}

The remaining possibility is either the ISW effect caused by the 
gravitational potential wells of the galaxy clusters hosting the SNe 
or some contaminating emission from dust and/or gas residing in the 
vicinity of the SN host galaxies. 
The observed frequency dependence of $T_{\rm SN}$ from $z_{\rm SN}$ 
and the fact that the strongest correlation corresponds to the Planck 
frequency channel of 143 GHz suggest that the spectral energy distribution 
of the anomalous emission is similar to that of the CMB. 
This supports the idea of the ISW effect behind the 
seen correlation. Alternatively, if this anomaly is related to 
the emission from intergalactic matter distributed across large volumes 
at distances corresponding to $z>0.3$, then 
this matter must be thermalised.

Among the Planck products, there are two foreground distribution
maps derived by the Planck team by using the data from different frequency 
channels:  the Low-Frequency Foreground (LFF) and the Thermal Dust
Foreground (TDF) component maps. We have checked these maps 
for the presence of the discussed correlation. 
The calculated slopes of the regression lines $T_{\rm SN}/z_{\rm SN}$ 
and the corresponding Pearson's correlation coefficients for
these two foreground maps are given in 6th and 7th entries of 
Table~\ref{slopes1}. The LFF map (normalised to the 30 GHz channel)
does not show any significant correlation, whereas the TDF map
(normalised to 353 GHz) reveals a negative correlation similar to 
that of the Planck 353 GHz frequency channel but with 
higher significance of about $4.2\sigma$. Since this map
can contain contributions from both local and remote 
dust, the negative slope of the   $T_{\rm SN} / z_{\rm SN}$
correlation favours the remote black-body emission being 
responsible for this correlation.

The last entry of Table~\ref{slopes1} gives the results of our 
correlation analysis of the Planck CMB Lensing Potential Map (CLPM) 
which reflects the distribution of gravitational potential wells 
leading to the ISW effect (since the CLPM data are dimensionless,
we have scaled them to achieve a slope amplitude similar to those 
of the other entries of Table~\ref{slopes1}). 
The very low significance of the correlation between the CLPM
amplitudes and $z_{\rm SN}$ casts the ISW effect to be a 
less favourable candidate to the mechanism responsible for  
the discussed $T_{\rm SN}/z_{\rm SN}$ correlation.

\section{Conclusions}

By using the recently released Planck satellite data we have 
not only confirmed the existence of the CMB anomaly previously reported 
for the WMAP data, but we have also found that the statistical significance of the 
correlation between the SNe redshifts and the temperature fluctuations of 
the CMB at the SNe locations is quite high. This significance
becomes higher if,  for our correlation analysis, 
we use a subsample of SNe of only the type Ia.

We have also shown that the found effect is frequency-dependent,
with this dependence being opposite to the SZ effect.
In our view, this correlation is likely to be related 
to either the integrated Sachs-Wolfe effect,
some remote foreground emission from dust and\,/or\,gas, 
or to a combination of both effects.   
  
If this anomaly is related to the 
clumps of matter (remote galaxies or galaxy clusters) traced by 
the presence of SNe then it might explain the 
findings of other authors showing a positive correlation 
between CMB and overdensities of galaxies  -- see, e.g., \citet{ho08}. 
The possible impact of dusty galaxy clusters 
on the estimates of the cosmological parameter was also discussed by 
 \citet{serra08}, \citet{dunkley10} and \citet{millea11}.

The observed systematic relationship between the SN redshifts  
and CMB temperature fluctuations should be further examined 
with the purpose to use it for reducing the contamination of 
CMB maps by distant foreground   
at $z > 0.5$ and, thus,  to increase the accuracy of 
determining the cosmological parameters.

\section*{acknowledgements}
\noindent
The research has made use of the following archives:
the Planck Legacy Archive 
operated by ESA's space astronomy centre;
the Hierarchical Equal Area iso-Latitude Pixelisation ({\scriptsize HEALPIX}) software
packages maintained by the Jet Propulsion Laboratory,
California Institute of Technology, Pasadena;
and the Supernovae Catalogue maintained by 
Sternberg Astronomical Institute, Moscow State University.
The authors are grateful to the anonymous Referee whose
very useful suggestions resulted in the essential improvements
of this paper.

\section*{Supporting Information}

Additional Supporting Information may be found in the online version of 
this article:

\noindent
{\bf Table 3}

\noindent
\url{http://mnras.oxfordjournals.org/lookup/suppl/doi:10.1093/mnras/stu1932/-/DC1}


\begin{thebibliography}{99}
{\small 


\bibitem[\protect\citeauthoryear{Bartunov et al.}{2007}]{bartunov07}
Bartunov O. S., Tsvetkov D. Yu., Pavlyuk N. N., 2007, Highlights in Astr.,
14, 316

\bibitem[\protect\citeauthoryear{Childress et al.}{2013}]{childress13}
Childress M., Aldering G., Antilogus P., et al., 2013, ApJ, 770, 107 


\bibitem[\protect\citeauthoryear{Di Paola et al.}{2002}]{dipaola02}
Di Paola A., Larionov V., Arkharov A., et al., 2002, A\&A, 393, L21

\bibitem[\protect\citeauthoryear{Dunkley et al.}{2011}]{dunkley10}
Dunkley J., et al., 2011, ApJ, 739, 52 

\bibitem[\protect\citeauthoryear{Efron}{1979}]{efron79}
Efron B., 1979, Ann. Statist., 7, 1

\bibitem[\protect\citeauthoryear{Elias-Rosa et al.}{2008}]{elias08}
Elias-Rosa N., Benetti S., Turatto M., et al., 2008, MNRAS, 384, 107

\bibitem[\protect\citeauthoryear{Farrah et al.}{2002}]{farrah02}
Farrah D., Meikle W. P. S., Clements D., et al., 2002, MNRAS, 336, L17

\bibitem[\protect\citeauthoryear{Filippenko}{1989}]{filippenko89}
Filippenko A. V., 1989, Publ. Astr. Soc. Pacific, 101, 588

\bibitem[\protect\citeauthoryear{Gorski et al.}{2005}]{gorski05}
Gorski K. M., Hivon E., Banday A. J., et al., 2005, ApJ, 622, 759

\bibitem[\protect\citeauthoryear{Goto et al.}{2012}]{goto12}
Goto T., Szapudi I., Granett B. R., 2012, MNRAS, 422, L77

\bibitem[\protect\citeauthoryear{Ho et al.}{2008}]{ho08}
Ho S., Hirata C. M., Padmanabhan N., et al., 2008, Phys. Rev. D, 78, 043518


\bibitem[\protect\citeauthoryear{Kozlovski et al.}{2010}]{kozlovski10}
Kozlovski S., Kochanek C. S., Stern D., et al., 2010, ApJ, 722, 1624

\bibitem[\protect\citeauthoryear{Mattila et al.}{2012}]{mattila12}
Mattila S., Dahlen T., Efstathiou A., et al., 2012, ApJ, 756, 111

\bibitem[\protect\citeauthoryear{Millea et al.}{2012}]{millea11}
Millea M., Dor\'e O., Dudley J., et al., 2012, ApJ, 746, 4 

%
%
\bibitem[\protect\citeauthoryear{Planck Collaboration}{2014}]{planck14}
Planck Collaboration,  2014, A\&A, in press, doi: 10.1051/00004-6361/201321529,  
arXiv:1303.5062[astro-ph.CO]

\bibitem[\protect\citeauthoryear{Romero-Canizales et al.}{2012}]{romero11}
Romero-Canizales C., Matilla S., Alberdi A., et al., 2011, MNRAS, 415, 2688

\bibitem[\protect\citeauthoryear{Sachs \& Wolfe}{1967}]{sachs67}
Sachs R. K., Wolfe A. M. 1967, ApJ, 147, 73

\bibitem[\protect\citeauthoryear{Serra et al.}{2008}]{serra08}
Serra P., Cooray A., Amblard A., 2008, Phys. Rev. D 78, 043004

\bibitem[\protect\citeauthoryear{Sunyaev \& Zeldovich}{1970}]{sunyaev70}
Sunyaev R. A., Zeldovich Ya. B., 1970, Astrophys. Space Sci., 7, 3

\bibitem[\protect\citeauthoryear{Vink}{2004}]{vink04}
Vink J., 2004, ApJ, 604, 693

\bibitem[\protect\citeauthoryear{Yershov et al.}{2012}]{yershov12}
Yershov V. N., Orlov  V. V., Raikov A. A., 
2012, MNRAS, 423, 2147

 }
\end{thebibliography}
\end{document}